\documentclass[twocolumn]{article}
\usepackage[dvips]{graphicx}

\author{R.\ Burghaus\\Institut f{\"u}r Theoretische Physik IV\\
  Heinrich-Heine-Universit{\"a}t D{\"u}sseldorf\\Universit{\"a}tsstr.~1\\
  D-40225 D{\"u}sseldorf, Germany}
\title{Super lattice formation of an array of volatile wetting droplets}
\date{}

\begin{document}
\sloppypar

\maketitle

\begin{abstract}
For an ordered array of critical volatile wetting droplets the
formation of a super lattice by an Ostwald-ripening like competition
process is considered. The underlying diffusion problem is treated
within a quasistatic approximation and to first order in the inverse
droplets distance.  The approach is rather general but a square
lattice and a triangular lattice are studied explicitly. Dispersion
relations for the super-lattice growth of these arrays are calculated.
\end{abstract}

In a recent experiment C.\ Sch{\"a}fle\ {\sl et al}\ \cite{Schaefle}
nucleated diethylene glycol droplets out of a supersaturated vapor
phase on a hexagonal array of hydrophilic (alkanethiol) patches on a
hydrophobic substrate. The pressure of the system was reduced
afterwards so that the droplets started to shrink by evaporation.
During this process the hexagonal lattice split into two triangular
super lattices in a sense that the droplets on one super lattice
shrank faster than those on the other super lattice. The super lattice
of faster shrinking droplets disappeared, finally. The remaining
triangular droplet lattice appeared to be stable against further
development of super structures.  The appearance of the triangular super
lattice from the hexagonal lattice is interpreted as an Ostwald-ripening
type process. The system reduces its free energy by having few large
droplets instead of many small
\cite{LifSly,Wagner,BurghausE96}. Similar observations where made by
the same authors\ \cite{Schaefle} as well as by Lacasta\ {\sl et al}\
\cite{Lacasta} simulating Cahn-Hilliard dynamics of droplet arrays in
a concentration field. The study of super lattice growth on an
array of wetting droplets is not only of fundamental interest as an
example of pattern formation and the dynamics of wetting on structured
surfaces, but it is also important to learn about the features of
volatile droplet arrays since they can be used to build e.g.\ chemical
sensors (see \cite{Kumar,Morhard}).

The scope of this paper is the analytic discussion of the diffusional
growth properties of a set of wetting droplets sitting on hydrophilic
patches on a hydrophobic substrate. We are considering circular patches
of equal radius $a$. Square and triangular lattice configurations of
the patches will be studied explicitly. It is assumed that there
is a wetting droplet sitting in the center of each patch. The
different possible equilibrium configurations and phase transitions in
such a system were discussed in great detail by Lenz and Lipowsky only
recently \cite{Lenz,LenzPhD}.

To discuss the diffusional growth of a wetting droplet out of the
surrounding gas phase, one has to specify the boundary conditions of
the diffusion equation on the substrate and on the droplets
surface. Since there is no diffusion flux into the substrate, the
normal derivative of the concentration field vanishes there. The
concentration $c_s$ in the gas phase on top of the droplet is given by
the Gibbs-Thomson relation. It is a constant along the droplets
surface and is just a function of the radius of curvature $R$
\cite{BurghausE96,cond-mat}
\begin{equation}
        c_s(R) = c_0 \left( 1 + {\Lambda \over R} \right),
        \label{GT}
\end{equation}
where $c_0$ is the concentration above a flat condensate and $\Lambda$
is a capillary length.

We assume that the droplet is growing slowly due to its high density
compared to the surrounding gas phase. The diffusion equation, which
describes the concentration around the droplet, can be approximated by
the Laplace-equation $\Delta c = 0$, therefore. The Neumann condition
of vanishing normal derivative on the substrate suggests to mirror the
system on the substrate to fulfill the condition implicitly. This way
the semi-infinite system becomes an infinite system formally and the
spherical-cap shaped wetting droplet becomes a symmetric lens that
provides a uniform (Dirichlet) boundary condition to the outside
concentration field. Since this field is described by a Laplace
equation, one can use an electrostatic analogy where the electrical
potential corresponds to the concentration field. The mirrored
droplet, i.e.\ the symmetric lens, is a conductor in this picture
\cite{BurghausE96,cond-mat,Picknett}.  The volume growth $\dot\Omega$
of an individual droplet corresponds to the charge of the analogous
conductor because the diffusion flux density on the droplets surface
is given by the normal derivative of the concentration field and the
charge density of the conductor is given by the electric field on top
of the conductor which is the derivative of the potential. This
argument leads to the growth equation
\begin{equation}
        \dot\Omega = C(R,\theta) \cdot \left(c_\infty - c_s(R)\right)
        \label{singledroplet}
\end{equation}
eventually \cite{BurghausE96,cond-mat}, where $C(R,\theta)$ is the
electric capacity of the symmetric lens and $c_\infty$ is the systems
concentration far away from the droplet.  The capacity $C(R,\theta)$
depends linearly on the radius of curvature $R$ and is a function of
the wetting droplets contact angle $\theta$. It can be found in the
literature \cite{BurghausE96,Picknett,Snow49}.

\begin{figure}[ht]
\begin{center}
\leavevmode
\vbox{
\includegraphics[scale=.8]{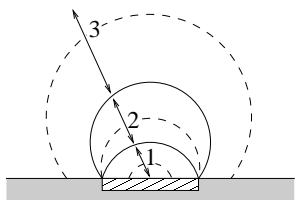}}
\end{center}
 \caption{Three qualitatively different droplet sizes can be
   distinguished: (1) A droplet which sits on the hydrophilic patch
   completely; (2) the contact line of the droplet sits on the edge of
   the hydrophilic region; (3) the contact line is on the hydrophobic
   substrate completely.}
  \label{fig:DropletRegimes}
\end{figure}

Since the surface tension between the droplet and the hydrophilic
patch is smaller than the one between the droplet and the hydrophobic
substrate, the contact (Youngs) angles are different in these
regions.  A small droplet, which completely sits on a hydrophilic
patch, has a small contact angle $\theta_1$. If the droplet is too big
to fit onto the patch it reaches out on the hydrophobic region where
its contact angle $\theta_2$ is bigger. Consequently one can
distinguish three different volume regimes for the droplets. Regime
(1), where the contact line is on the hydrophilic patch
completely, regime (2), where the contact line is fixed on the border
between the hydrophilic and the hydrophobic material and regime (3), where the
contact line is completely in the hydrophobic region (see
fig.~\ref{fig:DropletRegimes}).

In regime (2) droplet growth takes place with a fixed contact line but
a changing contact angle $\theta$. If $\theta < \pi/2$, this leads to
the unusual case that the radius of curvature $R$ decreases with an
increasing droplet volume. This implies, that in this case a droplet
in an environment $c_\infty = c_s(R)$ is \emph{not} critical but
stable because its Gibbs-Thomson boundary concentration increases with
increasing volume unlike the usual case where the Gibbs-Thomson
boundary condition decreases with increasing droplet size. This
stability allows to grow droplets of identical size on an array of
equally sized patches in the regime (2), which is important for
experimental purposes.

Now there is not only one droplet in the system but the collective
behavior of a system of droplets shall be studied. The boundary
condition on top of every single droplet is given by (\ref{GT}), so
that we can use the full electrostatic analogy again. We have a
system of conductors where the relation between the charges and the
potentials is given by a capacitance matrix which now yields
\begin{equation}
        \dot\Omega_i = \sum_j{C_{ij} \cdot \left(c_\infty -
        c_s(R_j)\right)}
        \label{manydroplet}
\end{equation}
for the growth rates of the individual droplets generalizing
eq.~(\ref{singledroplet}). Here the indexes $i,j$ count the individual
droplets, $\Omega_i$ and $R_i$ are the volume and the radius of
curvature of droplet $i$ and the matrix $C_{ij}$ is the abovementioned
capacitance matrix of the system.

In the following we will assume that the lattice constant $d$ of the
droplet array is large compared to the droplet sizes ($R_i \ll
d$). Then, the inverse of the capacitance matrix can be written
\begin{equation}
C^{-1}_{ij}=\left\{
\begin{array}{ll}
C_i^{-1}(R,\theta) + \mathcal{O}\left(1/{d^2}\right)&\textrm{for}\quad i
= j\\
1/R_{ij} + \mathcal{O}\left(1/{d^2}\right)&\textrm{for}\quad i \ne
j
\end{array}\right.
\label{Cinv}
\end{equation}
where $C_i(R,\theta)$ is the capacity of the droplet $i$
alone. $R_{ij}$ is the distance between the centers of the droplets
$i$ and $j$. Eqs.~(\ref{GT}), (\ref{manydroplet}) and (\ref{Cinv})
lead to
\begin{equation}
        c_0 \left( 1 + {\Lambda \over R_i} \right) = c_\infty + {1
        \over C_i(R_i,\theta_i)}\dot\Omega_i + \sum_{j \ne i}{1\over
        R_{ij}}\dot\Omega_j
        \label{growth}
\end{equation}
to linear order in inverse droplet distances.

If we now consider a system where each droplet satisfies $R_i = R_c :=
(\Lambda c_0) / (c_\infty - c_0)$ (i.e. it is critical if its radius
increases with increasing volume or stable if its radius decreases with
increasing volume), $\dot R_i = 0$ is a (in most cases unstable)
solution of the growth equation. To study the onset of the
spontaneous development of a super lattice on the regular lattice of
wetting droplets, we introduce $\delta R_i \cdot e^{\omega t}:= R_i -
R_c$ where the $\delta R_i$ describe an eigenmode of the linearized
version of eq.~(\ref{growth}). The growth exponent reads
\begin{equation}
        \omega = {-c_0 {\Lambda \over R_c^3} \over 2 \Omega^\prime
        \left[C_c + \sum_{j \ne i}{1 \over R_{ij}}{\delta R_j
        \over \delta R_i}\right]}.
        \label{lineargrowth}
\end{equation}
Here $C_c$ is the capacity of an individual critical droplet alone and
$\Omega^\prime := d\Omega(R,\theta(R)) / dR\,|_{R=R_c}$ describes the
variation of the droplet volume when its radius $R$ is changed obeying
the physical boundary conditions on the contact angle, i.e.
holding the contact angle constant if we are in the regime~1 or 3 of
fig.~\ref{fig:DropletRegimes} and pinning the contact line (varying
the contact angle) if we are in the regime 2. As mentioned earlier,
$\Omega^\prime$ is negative in regime 2 if $\theta < \pi/2$.

The sum $\Sigma := d\cdot\sum_{j \ne i}{(1 / R_{ij})}{(\delta R_j /
\delta R_i)}$ is independent of the choice of $i$ since we look at
eigenmodes of eq.~(\ref{growth}). It depends on the type of droplet
lattice and the mode under consideration. One finds a divergence at
$\Sigma/d = -C_c$. This divergence is an artifact of the approximation
for the capacitance matrix to first order in $1/R_{ij}$,
i.e. neglecting $\mathcal{O}(1/d^2)$. The fastest growing mode is
found at the smallest $\Sigma$. In the following the growth exponent
$\omega$ will be evaluated for the different modes of an infinite
square lattice and an infinite triangular lattice.

\begin{figure}[ht]
\begin{center}
\leavevmode
\vbox{
\includegraphics[scale=.6]{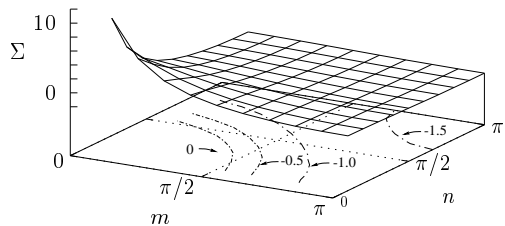}}
\end{center}
 \caption{The sum $\Sigma$ as a function of the wavenumbers $m$ and
  $n$ for a square lattice.}
  \label{fig:SigmaSquare}
\end{figure}

Due to the symmetry of the square lattice the eigenmodes of the growth
equation are just Fourier modes. In fig.~\ref{fig:SigmaSquare} the sum
$\Sigma$ is plotted as a function of $m$ and $n$, where $m$, $n$ are
the wavenumbers along the two axis of the square lattice. The minimum
is found at $\Sigma(\pi,\pi)$ which means that the preferred growth
mode leads to a checkerboard structure. The two corresponding square
super lattices on the original square lattice carry uniformly
shrinking or uniformly growing droplets, respectively. This behavior is
optimal in the sense that every growing droplet is surrounded only by
shrinking droplets and the other way around leading to a fast exchange
of matter.

\begin{figure}[ht]
\begin{center}
\leavevmode
\vbox{
\includegraphics[scale=.6]{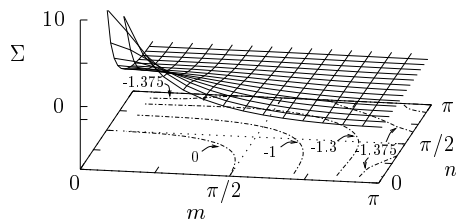}
}
\end{center}
 \caption{The sum $\Sigma$ as a function of the wavenumbers $m$ and
  $n$ for a triangular lattice.}
  \label{fig:SigmaTriangular}
\end{figure}

\begin{figure}[ht]
\begin{center}
\leavevmode
\vbox{
\includegraphics[scale=.3]{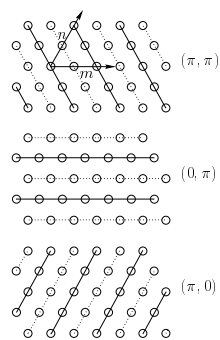}
}
\end{center}
 \caption{The fastest growing modes of a triangular lattice.}
  \label{fig:TriMode}
\end{figure}

In case of a triangular lattice things are not as obvious since the
geometry does not allow every droplet to be surrounded by droplets of
a different growth sign without frustration. If we introduce two non
rectangular axis along two of the primary axis of the triangular
lattice we have again translational symmetry along these axis and can
Fourier analyze the system. We get a sum $\Sigma$ for the
corresponding wavenumbers as shown in
fig.~\ref{fig:SigmaTriangular}. The dispersion relation has a
threefold degenerate minimum at $(\pi,\pi)$, $(0,\pi)$ and $(\pi,0)$
reflecting the special symmetry of the triangular lattice as shown in
fig.~\ref{fig:TriMode}. These fastest eigenmodes can be characterized
by the property that each droplet has four nearest neighbors with opposite
growth signs and two nearest neighbors with the same growth sign. In
this respect the growth conditions in the triangular lattice are
inferior to those in the square lattice. This is reflected by the fact
that the minimum $\Sigma$ of the square lattice is lower than the one
of the triangular lattice, i.e. the super lattice formation is faster
in the square lattice case..

C.\ Sch\"afle et al. found in their experiment \cite{Schaefle} that
the triangular droplet lattice that occurred as a super lattice of the
initial hexagonal lattice did not show any further instability,
whereas the calculations provided in this paper display an instability
as sketched in fig.~\ref{fig:TriMode}. Unlike the subject of this
paper the experiment does not analyze critical droplets so that in
addition to $\omega$ there is a timescale related to the overall
evaporation of droplets. It seems reasonable to expect that the
instability of the triangular droplet lattice could be seen
experimentally by stretching the timescale provided by the overall
evaporation, i.e. by going closer to the critical concentration.

I like to thank C.\ Sch\"afle \emph{et al.} for sharing their
experimental results with me prior to publication. I'm especially
grateful to B.\ Schmittmann and R.\ K.\ P.\ Zia for their kind
hospitality and helpful discussions. This work is supported by the
Deutsche Forschungsgemeinschaft via BU 1172/1-1 and ``Benetzung und
Strukturbildung an Grenzfl\"achen'' as well as the EU via FMRX-CT
98-0171 ``Foam Stability and Wetting Transitions''.

% Figures

\begin{thebibliography}{99}
\bibitem{Schaefle} C. Sch{\"a}fle {\sl et al.} (to appear in Phys.~Rev.~Lett.)
\bibitem{LifSly} I.~M.~Lifshitz, V.~V.~Slyozov,
  J.~Phys.~Chem.~Solids {\bf 19}, 35 (1961)
\bibitem{Wagner} C.~Wagner, Z.~Elektrochem.~{\bf 65}, 581 (1961)
\bibitem{BurghausE96} R.~Burghaus, Phys.~Rev.~{\bf E} 54, 6955
  (1996)
\bibitem{Lacasta} A. M. Lacasta, I. M. Sokolov, J. M. Sancho and
F. Sagu{\'e}s, Phys. Rev. E, {\bf 57}, 6198 (1998)
\bibitem{Kumar} A. Kumar {\sl et al.}, Acc. Chem. Res. {\bf 28}, 219 (1995)
\bibitem{Morhard} F. Morhard {\sl et al.}, Proc. ECS {\bf 97-19}, 1058 (1997)
\bibitem{Lenz} P.~Lenz and R.~Lipowsky, Phys.~Rev.~Lett.~{\bf 80},
1920 (1998)
\bibitem{LenzPhD} P. Lenz, ``Benetzungsph\"anomene auf strukturierten
Substraten'', PhD-thesis (1998)
\bibitem{cond-mat} R.\ Burghaus cond-mat/9903107
\bibitem{Picknett}R.~G.~Picknett, R.~Bexon, J.~Coll.~Interf.~Sci.~{\bf
  61}, 336 (1977)
\bibitem{Snow49}C.~Snow, J.~of Research~NBS, {\bf 43}, 377 (1949)
\end{thebibliography}
\end{document}